\documentclass[reprint,prd,showpacs,nofootinbib]{revtex4-1}

\usepackage{graphicx}
\usepackage{latexsym,amsmath,amssymb,lmodern,float,url}
\usepackage{natbib}
\usepackage[bookmarks,linktocpage,pdfpagelabels,plainpages=false,hyperfigures,linkcolor=blue,citecolor=blue]{hyperref} 
\hypersetup{colorlinks=true}
\usepackage{scrextend}
\usepackage[percent]{overpic}
\newcommand{\elll}{\ell^+\ell^-}
\newcommand{\mm}{(\mu^+\mu^-)}
\newcommand{\eep}{\eta/\eta'}
\newcommand{\br}{\mathcal{BR}}

\begin{document}
\title{Scouring meson decays for true muonium}

\author{Yao Ji}
\email{yao.ji@physik.uni-regensburg.de}
\affiliation{Institut f\"ur Theoretische Physik, Universit\"at Regensburg, Regensburg 93040, Germany}
\author{Henry Lamm}
\email{hlamm@umd.edu}
\affiliation{Department of Physics, University of Maryland, College Park, MD 20742}
\date{\today}

\begin{abstract}
Rare meson decay experiments promise to measure branching ratios as small as $10^{-13}$.  This presents an opportunity to discover the $\mu^+\mu^-$ bound state true muonium.  We consider a set of possible channels, all with branching ratios above $\sim10^{-11}$.  For the electromagnetic decays $\eta/\eta'\rightarrow(\mu^+\mu^-)\gamma$, theoretical and phenomenological form factors $F_{\eta/\eta'\gamma\gamma^*}(Q^2)$ allow predictions of $\mathcal{BR}(\eta'\rightarrow \mm \gamma)\sim4.8\times10^{-10}$ and $\mathcal{BR}(\eta'\rightarrow \mm \gamma)\sim3.7\times10^{-11}$ at the 5\% level.  Discussion of experimental prospects and potential backgrounds are made.
\end{abstract}

\maketitle

Within the Standard Model, only the Higgs interaction breaks {\it lepton universality\/}, but the discovery of neutrino masses implies that at least one beyond-Standard Model modification is required.  Many precision physics searches have been undertaken in the charged lepton sector to detect additional lepton universality violations. Measurements of $(g-2)_\ell$~\cite{PhysRevD.73.072003}, charge radii~\cite{Antognini:1900ns,Pohl1:2016xoo}, and $B$ meson decays~\cite{Aaij:2014ora,Lees:2012xj,Lees:2013uzd,Huschle:2015rga,
Sato:2016svk,Aaij:2015yra,Hirose:2016wfn,Aaij:2017uff,Aaij:2017deq,Hirose:2017dxl,Aaij:2017tyk} have each shown hints of discrepancy.  The bound state of $\mm$, \textit{true muonium}, or TM for short, presents another avenue for investigating lepton universality~\cite{TuckerSmith:2010ra,Lamm:2016jim}.  To facilitate these studies, efforts are on-going to improve theoretical predictions~\cite{Jentschura:1997tv,Jentschura:1997ma,PhysRevD.91.073008,Lamm:2016vtf,PhysRevA.94.032507,Lamm:2017lib,Lamm:2017ioi}.  Alas, true muonium remains undetected today.

There are two categories of $\mm$ production methods discussed within the literature: particle collisions (fixed-target and collider)~\cite{Bilenky:1969zd,Hughes:1971,Moffat:1975uw,Holvik:1986ty,Ginzburg:1998df,ArteagaRomero:2000yh,Brodsky:2009gx,Chen:2012ci}, or through rare decays of mesons~\cite{Nemenov:1972ph,Vysotsky:1979nv,Kozlov:1987ey,Malenfant:1987tm,Martynenko:1998jk,Ji:2017lyh,Fael:2018ktm}.  Both are challenging due to the low production rates.  Currently, the HPS~\cite{Celentano:2014wya} experiment is searching for true muonium~\cite{Banburski:2012tk} via $e^-Z\rightarrow\mm X$. Another fixed-target experiment, DIRAC~\cite{Benelli:2012bw} could look for $\mm$ in an upgraded run~\cite{dirac}.  The existing KOTO experiment~\cite{Ahn:2016kja} and proposed NA62-KLEVER~\cite{Moulson:2016zsl} hope to achieve sensitivities to $K_L$ decays with $\mathcal{BR}\sim10^{-13}$, which would also present an opportunity to detect true muonium~\cite{Ji:2017lyh}.

In this work, we present predictions for $\mathcal{BR}(\eta/\eta'\rightarrow\mm \gamma)$ where true muonium is accompanied by a monochromatic photon (which in the $\eep$ rest frame are 233.2 MeV and 455.6 MeV respectively) including $\mathcal{O}(\alpha)$ radiative corrections.  These calculations improve upon previous constituent quark model calculations which estimated $\mathcal{BR}(\eta\rightarrow\mm \gamma)\approx10^{-9}$~\cite{Nemenov:1972ph,Kozlov:1987ey}. Numerous studies of $F_{\eta/\eta'\gamma\gamma*}(Q^2)$ are available to estimate potential systematics for this process.  Other discovery channels involving hadronic final states with $\mathcal{BR}\geq10^{-12}$ are discussed and comments on backgrounds are made.

The order of magnitude of $\mathcal{BR}(X\rightarrow Y (\elll))$ can be estimated by multiplying $\mathcal{BR}(X\rightarrow Y \gamma)$ by $\alpha^4$, which arises from $|\mathcal{M}(\gamma^*\rightarrow\elll)\mathcal{M}(\elll\rightarrow (\elll))|^2\propto \alpha|\psi(0)|^2$.  This implies that $\mathcal{BR}(X\rightarrow Y \mm)\lessapprox10^{-9}$.  In Table~\ref{tab:brz}, we have included a list of channels which led to branching ratios to true muonium of $\gtrapprox10^{-12}$.  There are two processes with $\mathcal{BR}>10\%$: $\eta'\rightarrow \rho \gamma$ and the electromagnetic $\eta\rightarrow \gamma\gamma$, the latter along side the percent level $\eta'\rightarrow \gamma\gamma$ we will discuss in more detail.  The large number of $J/\psi$ events being collected in the near-future also presents the possibility of subpercent channels like $J/\psi\rightarrow \eta'\gamma$, as well as the percent-level but more complicated inclusive $J/\psi\rightarrow X_h\gamma$~\cite{Besson:2008pr}.  Predictions of $X_h+\mm$ decays require knowledge of hadronic transition form factors, and are more limited in precision compared to processes like $\eep\rightarrow \mm\gamma$ where the better studied electromagnetic form factors are available.  These electromagnetic decays are the main focus of the rest of the paper. 

\begin{table}
 \caption{\label{tab:brz}Meson decay branching ratios involving photons considered in this work.  The first two are electromagnetic decays, while the others are strong decays.  Branching ratios to true muonium can be estimated by multiplying by $\alpha^4\approx2.8\times10^{-9}$}
 \begin{center}
 \begin{tabular}{l c}
 \hline\hline
 Channel & $\mathcal{BR}$\\
\hline
$\eta\rightarrow \gamma\gamma$&$4.0\times10^{-1}$\\
$\eta'\rightarrow \gamma\gamma$&$2.2\times10^{-2}$\\
\hline
$\eta'\rightarrow \rho \gamma$&$2.9\times10^{-1}$\\
$\omega\rightarrow \pi^0\gamma$&$8.2\times10^{-2}$\\
$\eta'\rightarrow \omega\gamma$&$2.8\times10^{-2}$\\
$J/\psi\rightarrow \eta'\gamma$ & $5.2\times10^{-3}$\\
$J/\psi\rightarrow X_h\gamma$ & $6\times10^{-2}$\\
\hline\hline
 \end{tabular}
\end{center}
\end{table}

While $K_L$ experiments are reaching sensitivities of $10^{-13}$, the sensitivity of $J/\psi$ and $\eep$ searches is worse. At present, the BESIII experiments has the largest $J/\psi$ and $\eep$ data sets. From the $1.3\times10^{9}$ $J/\psi$ events collected and using the two largest branching ratios, $\gamma\eep$ and $\phi\eep$, one anticipates $2.0\times10^6$ $\eta$ events and $7.7\times10^6$ $\eta'$ events with a factor to 10 increase expected in the next decade~\cite{Fang:2017qgz}.  This $10^{10}$ $J/\psi$ is right at the edge of what is necessary for an inclusive search.  Another similar sized data set exists from A2 where $\eta$ is produced through $\gamma p\rightarrow \eta p$, and have $6.2\times10^{6}$ $\eta$ events.  The recently approved JLab Eta Factory experiment anticipates collecting $1.3\times10^{8}$ $\eta$ and $9.8\times10^{7}$ $\eta'$ events with 200 days of beam time~\cite{ganjlab}, and beyond that can be run in parallel with the GlueX experiment if the later is extended beyond 2023~\cite{ganjlab}.  Further into the future, proposals like REDTOP at FermiLab discussed methods to accrue $10^{13}\,\eta$ and $10^{11}\,\eta'$ events~\cite{Gatto:2016rae}.

Following previous calculations for the electromagnetic decay of mesons to atoms~\cite{Nemenov:1972ph,Vysotsky:1979nv,Kozlov:1987ey,Malenfant:1987tm,Martynenko:1998jk,Ji:2017lyh}, the branching ratios are
\begin{align}
\label{eq:br1}
 \frac{\br(\eep\rightarrow\mm\gamma)}{\br(\eep\rightarrow\gamma\gamma)}=&\nonumber\\\frac{\alpha^4\zeta(3)}{2}(1-z_{TM})^3\bigg[1+&C_{\eep}\frac{\alpha}{\pi}\bigg]\left|f_{\eep}(z_{TM})\right|^2\, ,
\end{align}
where $\zeta(3)=\sum_n 1/n^3$ arising from the sum over all allowed $\mm$ states, $C_\eta=-0.35(6)$ and $C_{\eta'}=0.35(3)$, $z_{TM}=M_{TM}^2/M_{\eep}^2\approx4M_{\mu}^2/M^2_{\eep}$, and $f(z)=F_{\eep\gamma\gamma^*}(z)/F_{\eep\gamma\gamma^*}(0)$.  

In \cite{Ji:2017lyh}, the $\mathcal{O}(\alpha)$ radiative corrections to the analogous process $K_L\rightarrow \mm \gamma$ including the leptonic and hadronic vacuum polarization~\cite{PhysRevA.94.032507}, and an improved calculation of the double virtual photon contribution $h \rightarrow \gamma^*(k)+\gamma^*(P_{h}-k)\rightarrow\gamma + \text{TM}$ were presented, where $P_{h}$ is the four-momentum of the initial hadron, and $k$ is the four-momentum of one of the virtual photons.  This final term is dependent upon the initial state and has been recomputed for the $\eep$ processes.  

 \begin{figure}[!t]
\includegraphics[width=\linewidth]{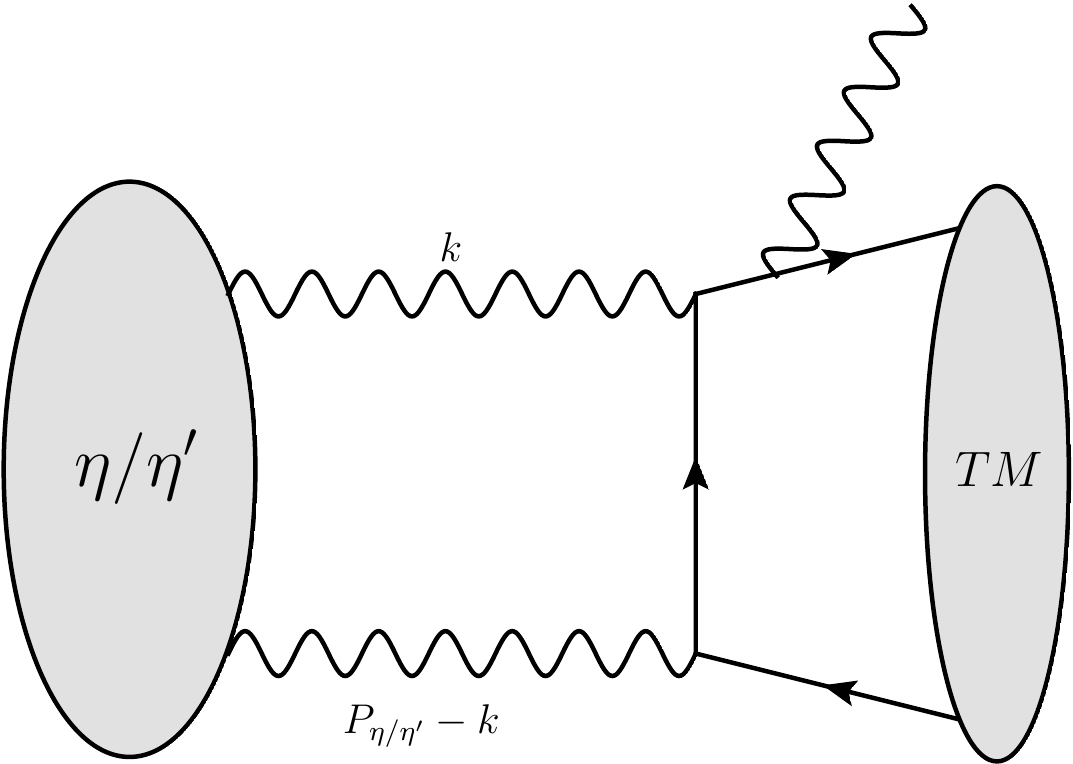}
\caption{One of the Feynman diagrams of $\eta/\eta'\rightarrow\gamma^*\gamma^*\rightarrow\mm\gamma$ which contribute to the branching ratio at ${\cal O}(\alpha^5)$ and is proportional to $F_{\gamma^*\gamma^*}(z_1,z_2)$}
\label{fig:Oalpha-additonal}
\end{figure}

For this contribution, one should take the convolution of the QED amplitude with double-virtual-photon form factor $F_{\eta/\eta'\gamma^*\gamma^*}(k^2/M_{\eta/\eta'}^2,(P_{\eta/\eta'}-k)^2/M_{\eta/\eta'}^2)$. Taking the form factor to be a constant equal to $F_{\gamma\gamma^*}(0,z_{TM})$ and factoring it out from the integral is a sufficient approximation as shown in~\cite{Kampf:2005tz}.  This approximation is expected to receive process-dependent corrections but all comparable processes have errors less than $10\%$~\cite{Kampf:2018wau, Husek:2015wta} with $4\%$ being a reasonable estimate for our particular process based on various models for the form factors.  With this, we find that the total next to leading order correction is
\begin{align}
 C_{\eep}=&[C_{e VP}+C_{\mu VP}+C_{\tau VP}+C_{hVP}+C_{ver}]+C_{\gamma^*\gamma^*}\nonumber\\
 =&\left\{
 \begin{aligned}
 &\left[\frac{8.526(4)}{9}\right]-\frac{11.7(5)}{9}\phantom{X}\text{for $\eta$}\\
 &\left[\frac{8.526(4)}{9}\right]-\frac{5.4(3)}{9}\phantom{X}\,\,\,\text{for $\eta'$}
 \end{aligned}
 \right.\, ,
\end{align}  
where the bracketed terms are independent of the initial meson, the $C_{iVP}$ indicate vacuum polarization contributions from $i=e,\mu,$ and hadrons, $C_{ver}$ is the vertex correction term of~\cite{Vysotsky:1979nv}, while $C_{\gamma^*\gamma^*}$ is the contribution from diagram in Fig.~\ref{fig:Oalpha-additonal} and alike.  A similar calculation for positronium, where other lepton flavors and hadronic loop corrections are negligible, finds the $\displaystyle\frac{\alpha}{\pi}$ coefficient to be $C_0=C_{VP}+C_{ver}=-52/9$~\cite{Vysotsky:1979nv}.  $C_{iVP}$ are found by computing
\begin{equation}
 C_{iVP}=4m_\mu^2\int_{4m_i^2}^{\infty}\mathrm{d}t \frac{{\rm Im }\, \Pi (t)}{t(4m_\mu^2-t)}
\end{equation}
from the spectral functions ${\rm Im}\, \Pi(t)$.  This function is known to leading order analytically for the leptons, and is derived from experiment for the hadronic distribution.

$F_{\eep\gamma\gamma^*}(0)$ are fixed to the experimental values~\cite{Tanabashi:2018oca}
\begin{align}
 \br(\eta\rightarrow\gamma\gamma)=&39.41(20)\%\,,\nonumber\\
  \br(\eta'\rightarrow\gamma\gamma)=&2.22(8)\%. 
\end{align}

Evaluating Eq.~(\ref{eq:br1}), we find
\begin{align}
 \br(\eta\rightarrow\mm\gamma)=&4.14(3)\times10^{-10}|f\left(z_{TM}\right)|^2\, ,\nonumber\\
  \br(\eta'\rightarrow\mm\gamma)=&3.26(12)\times10^{-11}|f\left(z_{TM}\right)|^2\, ,
\end{align}
 where the dominant error is from $\mathcal{BR}(\eep\rightarrow\gamma\gamma)$, preventing the measurement of these radiative corrections from this ratio. An improved value of $\mathcal{BR}(\eep\rightarrow\gamma\gamma)$ or constructing a different ratio, as we do below, can allow sensitivity to these corrections.

 \begin{table*}[ht]
 \caption{\label{tab:1}Form factor coefficients and $\mathcal{BR}(\eep\rightarrow\mm\gamma)$.}
 \begin{center}
 \begin{tabular}{l c c c c c c c}
 \hline\hline
  &\multicolumn{2}{c}{$\eta$ Coefficients$\phantom{xxxx}$}&$\mathcal{BR}(\eta\rightarrow \mm\gamma)\times10^{10}$&\multicolumn{2}{c}{$\eta'$ Coefficients$\phantom{xxxx}$}&$\mathcal{BR}(\eta'\rightarrow \mm\gamma)\times10^{11}$&Ref.\\
  \hline
  $\chi$PT			&$b_{\eta}$&0.51		&4.79(3)	&$b_{\eta'}$&1.47	&3.74(14)	&\cite{Ametller:1991jv}\\
  VMD				&$b_{\eta}$&0.53		&4.82(3)	&$b_{\eta'}$&1.33	&3.70(14)	&\cite{Ametller:1991jv}\\
  CQ Loops	        &$b_{\eta}$&0.51		&4.79(3)	&$b_{\eta'}$&1.30	&3.69(14)	&\cite{Ametller:1991jv}\\
  BL Interp.	    &$b_{\eta}$&0.36		&4.59(3)	&$b_{\eta'}$&2.11	&3.96(15)	&\cite{Brodsky:1981rp}\\
  R$\chi$T - 1 Octet	&$b_{\eta}$&0.546(9)	&4.84(3)&$b_{\eta'}$&1.384(3)	&3.71(14)		&\cite{Czyz:2012nq}\\
  R$\chi$T - 2 Octets	&$b_{\eta}$&0.521(2)	&4.81(3)&$b_{\eta'}$&1.384(3)	&3.71(14)		&\cite{Czyz:2012nq}\\
  Anomaly SR		&$b_{\eta}$&0.51		&4.79(3)&$b_{\eta'}$&1.06	&3.61(13)		&\cite{Klopot:2013laa}\\
  Anomaly SR		&$b_{\eta}$&0.54		&4.83(3)&$b_{\eta'}$&1.16	&3.64(14)		&\cite{Klopot:2013laa}\\
  \hline
  CELLO				&$b_{\eta}$&0.428(89)	&4.68(12)	&$b_{\eta'}$&1.46(23)	&3.74(16)	&\cite{Behrend:1990sr}\\
  CLEO				&$b_{\eta}$&0.501(38)	&4.78(6)	&$b_{\eta'}$&1.24(8)	&3.67(14)	&\cite{Gronberg:1997fj}\\
  Lepton-G			&$b_{\eta}$&0.57(12)	&4.87(17)	&$b_{\eta'}$&1.6(4)	&3.79(20)	&\cite{Dzhelyadin:1980kh,Landsberg:1986fd}\\
  NA60				&$b_{\eta}$&0.585(51)	&4.89(8)	&--&--&--&\cite{Arnaldi:2009aa}\\
  WASA				&$b_{\eta}$&0.68(26)	&5.0(4)	&--&--&--&\cite{Hodana:2012rc}\\
  A2				&$b_{\eta}$&0.59(5)	&4.90(8)	&--&--&--&\cite{Aguar-Bartolome:2013vpw}\\
  \hline
  DA		        &$b_{\eta}$&$0.62^{+0.07}_{-0.03}$	&$4.94(10)$	&$b_{\eta'}$&$1.45^{+0.17}_{-0.12}$&3.74(15)	&\cite{Hanhart:2013vba}\\
  DA	            &$b_{\eta}$&$0.57^{+0.06}_{-0.03}$	&$4.87(9)$	&--&--&--&\cite{Kubis:2015sga}\\
  RA		        &$b_{\eta}$&$0.576(11)_{\rm st}(4)_{\rm sy}$	&&$b_{\eta'}$&$1.31(4)_{\rm st}(1)_{\rm sy}$	&&\\&$c_{\eta}$&$0.339(15)_{\rm st}(5)_{\rm sy}$&&$c_{\eta'}$&$1.74(9)_{\rm st}(3)_{\rm sy}$	&&\\&$d_{\eta}$&$0.200(14)_{\rm st}(18)_{\rm sy}$	&$4.953(30)_{\rm st}(6)_{\rm sy}$&$d_{\eta'}$&$2.30(19)_{\rm st}(21)_{\rm sy}$&$3.720(140)_{\rm st}(4)_{\rm sy}$		&\cite{Escribano:2015nra,Escribano:2015yup}\\
  \hline\hline
 \end{tabular}
\end{center}
\end{table*}

Results for the form factors $F_{\eep\gamma^*\gamma^*}(Q^2)$ can be broadly classified into three groups: theoretical predictions~\cite{Ametller:1991jv,Brodsky:1981rp,Czyz:2012nq,Klopot:2013laa}, experimental extractions~\cite{Behrend:1990sr,Gronberg:1997fj,Dzhelyadin:1980kh,Arnaldi:2009aa,Hodana:2012rc,Aguar-Bartolome:2013vpw}, and dispersion analyses~\cite{Hanhart:2013vba,Kubis:2015sga,Escribano:2015nra,Escribano:2015yup}.
The standard parameterization for $f(z)$ is a series expansion in $z$
\begin{equation}
\label{eq:ff}
 f(z)=1+b_{\eep}z+c_{\eep}z^2+d_{\eep}z^3+\cdots
\end{equation}
which for all but the dispersion analyses of~\cite{Escribano:2015nra,Escribano:2015yup} truncate at first order.  The theoretical predictions make vastly different assumptions about the coupling between $\eta/\eta'$ and the photon, as well as different modeling of the mixing between the two mesons.  The experimental results rely upon integrating the functional form of Eq.~(\ref{eq:ff}) in $Q^2$ bins, and then non-linearly fitting $b_{\eep}$.  The dispersive analyses rely upon connecting experimental data for multiple processes to the virtual photon form factors through analyticity and crossing symmetry.  We have tabulated all of the form factors considered in this work in Table~\ref{tab:1}. With the exception of the Brodsky-Lepage interpolation predictions of $\eta$~\cite{Brodsky:1981rp}, the branching ratios predicted for true muonium agree within the uncertainties.

While improving $\mathcal{BR}(\eep\rightarrow\gamma\gamma)$ is certainly desirable, one could consider other branching ratios that remove this uncertainty.  One potentially interesting ratio that would also test lepton universality is the ratio of true muonium to positronium $(e^+e^-)$:
 \begin{align}
R=&\frac{\mathcal{BR}(\eep\rightarrow\mm\gamma)}{\mathcal{BR}(\eep\rightarrow(e^+e^-)\gamma)}\nonumber\\
=&\frac{(1-z_{TM})^3\left(1+C_{\eep}\frac{\alpha}{\pi}\right)|f(z_{TM})|^2}{(1-z_{Ps})^3\left(1-\frac{52}{9}\frac{\alpha}{\pi}\right)|f(z_{Ps})|^2}\nonumber\\=&K_{\eep}\bigg|\frac{f(z_{TM})}{f(z_{Ps})}\bigg|^2,
\end{align}
which is independent of the $\mathcal{BR}(\eep\rightarrow\gamma\gamma)$ uncertainty and $K_\eta=0.62469(8)$ and $K_{\eta'}=0.87340(5)$. This ratio has the added feature that it directly measures lepton universality, and due to the small range of $Q^2$ of leptonic atom production, this ratio has reduced $Q^2$ dependence arising from form factor uncertainties.

Compared to this theoretical precision, the current and near-future experimental outlooks are less optimistic.  Clearly the present 10$^6\,\eep$ BESIII and A2 data  are insufficient for observing true muonium.  From these, one would expect to place a upper limits on the branching ratios ${\mathcal BR}\lesssim 10^{-5}$ which is $10^4-10^5$ times larger than the predicted rates.  This should be compared with the situation for $K_L\rightarrow \mm \gamma$ where upper bounds of ${\mathcal BR}\lesssim 10^{-9}$ are possible at KTEV~\cite{AlaviHarati:2001wd,Abouzaid:2007cm} which are superior limits but still $10^3$ times as large as predicted~\cite{Ji:2017lyh}.  In the next decade, BESIII's larger data set is still inadequate for even single-event detection through the $\eep$ processes, although the inclusive $J/\psi$ channel is potentially viable.  The JLab Eta Factory experiment would be competitive with the possible bounds from KTEV based on the Standard Model predictions.  What is required is a proposal like REDTOP at FermiLab which would be sufficient for a discovery of true muonium through the decay of $\eta$ with 100s of events, and potentially an observation of the $\eta'$ decay.  

The most promising signal channel for discovering true muonium in mesonic decays is $e^+e^-$, with a large background from the free decays $\eep\rightarrow \ell^+\ell^-\gamma$.  This background can be computed by integrating the differential cross section in an invariant mass bin, $M_{bin}$, centered around the $\mm$ peak defined as $[2m_\mu-M_{bin}/2,2m_\mu+M_{bin}/2]$.  For bin sizes similar to BESIII (20 MeV), the values are $\mathcal{BR}(\eta\rightarrow e^+e^-\gamma)_{bin}=4\times10^{-6}M_{bin}$, and $\mathcal{BR}(\eta'\rightarrow e^+e^-\gamma)_{bin}=3\times10^{-7}M_{bin}$ where $M_{bin}$ is in MeV.  This large raw background ($\sim10^{4}\times$ the signal) must be reduced, by its distinct features compared to true muonium decays can plausible do this.

The two two-body decay topology suggests cuts on momenta and angular distribution would be powerful in background suppression, but exact values of the suppression will be highly detector efficiency dependent.  As an example, for radiative Dalitz decay the angle $\theta_e$ between the $e^+e^-$ can be arbitrary, but from the true muonium decay $\theta_e\sim m_{TM}/E_{TM}\sim50^o\times\frac{\text{GeV}}{E_{\eep}}$. In BESIII, where the typically $\eep$ is produced from the decay of $J/\Psi$, one finds $\theta_e\sim30^o$.  This correlation can be more precisely measured than the invariant mass, and can yield a factor of 10 in background suppression.  Full reconstruction of the $\eep$ allows for cuts on the $\gamma$ energy, where the bin resolution is $\mathcal{O}(M_{bin})$ together with the anti-parallel correlation between the $\gamma$ and the true muonium yields at least factor of 3 further background suppression.  If the vertex resolution is better than 0.5 mm, cuts can be made using the proper lifetime of true muonium ground state $c\tau=0.5$ mm.  Otherwise nearly all the $e^+e^-$ will be insufficiently separated from the primary $\eep\rightarrow\mm\gamma$ vertex to distinguish.

\begin{acknowledgments}
HL is supported by the U.S. Department of Energy under Contract No. DE-FG02-93ER-40762.  YJ acknowledges the Deutsche Forschungsgemeinschaft for support under grant BR 2021/7-1.
\end{acknowledgments}
\bibliographystyle{apsrev4-1}
\bibliography{/home/hlamm/biblo}
\end{document}